\documentclass[aip,amsmath,amssymb,reprint,nofootinbib,floatfix]{revtex4-1}

\usepackage{graphicx} 
\usepackage{dcolumn}
\usepackage{bm}

\usepackage[utf8]{inputenc}
\usepackage[T1]{fontenc}
\usepackage{mathptmx}
\usepackage{etoolbox}

\begin{document}

\title[Electron Temperature Relaxation]{Electron Temperature Relaxation \\
      in the Clusterized Ultracold Plasmas}

\author{Yurii V. Dumin}
\email[Corresponding author's electronic mail: \\ ]{dumin@pks.mpg.de,
dumin@yahoo.com}
\affiliation{Max Planck Institute for the Physics of Complex Systems, \\
Noethnitzer Str.\ 38, 01187 Dresden, Germany}
\affiliation{Lomonosov Moscow State University,
Sternberg Astronomical Institute, \\
Universitetskii prosp.\ 13, 119234 Moscow, Russia}
\author{Anastasiia T. Lukashenko}
\email[Electronic mail: ]{lukashenko@dec1.sinp.msu.ru, a_lu@mail.ru}
\affiliation{Lomonosov Moscow State University,
Skobeltsyn Institute of Nuclear Physics, \\
Leninskie gory, GSP-1, 119991 Moscow, Russia}

\date{16 September 2022}

\begin{abstract}
Ultracold plasmas are a promising candidate for the creation of
strongly-coupled Coulomb systems.
Unfortunately, the values of the coupling parameter~$ {\Gamma}_e $ actually
achieved after photoionization of the neutral atoms remain relatively small
because of the considerable intrinsic heating of the electrons.
A conceivable way to get around this obstacle might be to utilize
a spontaneous ionization of the ultracold Rydberg gas, where the initial
kinetic energies could be much less.
However, the spontaneous avalanche ionization will result in a very
inhomogeneous distribution (clusterization) of the ions, which can change
the efficiency of the electron relaxation in the vicinity of such clusters
substantially.
In the present work, this hypothesis is tested by an extensive set of
numerical simulations.
As a result, it is found that despite a less initial kinetic energy, the
subsequent relaxation of the electron velocities in the clusterized plasmas
proceeds much more violently than in the case of the statistically-uniform
ionic distribution.
The electron temperature, firstly, experiences a sharp initial jump
(presumably, caused by the ``virialization'' of energies of the charged
particles) and, secondly, exhibits a gradual subsequent increase
(presumably, associated with a multi-particle recombination of the electrons
at the ionic clusters).
As a possible tool to reduce the anomalous temperature increase, we considered
also a two-step plasma formation, involving the blockaded Rydberg states.
This leads to a suppression of the clusterization due to a quasi-regular
distribution of ions.
In such a case, according to the numerical simulations, the subsequent
evolution of the electron temperature proceeds more gently, approximately
with the same rate as in the statistically-uniform ionic distribution.
\end{abstract}


\maketitle

\section{Introduction}
\label{sec:Intro}

Experimental realization of the Coulomb systems with large values of
the coupling parameter
\begin{equation}
\Gamma = \frac{q^2 n^{1/3}}{k_{\rm B} T}
\end{equation}
(where $ q $, $ n $, and $ T $ are the electric charge, concentration, and
temperature, respectively) is a long-standing problem in plasma
physics~\cite{Ichimaru82,Mayorov94,Killian07a}.
It was usually addressed~\cite{Fortov00} either by shock compression of
the substance, resulting in the increased concentration of the charged
particles~$ n $, or by employing the dusty particles with large effective
charges~$ q $.

An absolutely new way for the creation of strongly-coupled plasmas is
usage of the very low temperatures~$ T $, which became possible in the late
1990's and early 2000's due to advances in the laser cooling of atoms
in the magneto-optical
traps~\cite{Killian99,Gould01,Bergeson03,Killian07b}.
These devices were initially constructed for the creation of atomic
Bose--Einstein condensates, but later some of them were used for producing
the cold ionized gases and studying the respective plasma effects.
The main idea~\cite{Killian99} was that, by choosing the energy of
a narrow-band ionizing laser slightly above the ionization threshold of cold
(almost immobile) atoms, it would be possible to obtain a system of charged
particles with very low kinetic energy and, therefore, considerable values of
the coupling parameter~$ \Gamma $.

Besides, the ultracold plasmas could also be formed in the gas-dynamic
installations, \textit{e.g.}, the supersonic gaseous jets irradiated by
the laser beams~\cite{Morrison08}.
Such experiments benefit from the much greater gas density achieved,
but working with the molecular gases involves a lot of additional
complications~\cite{Aghigh20}.
Yet another (and even earlier) setup for the production of ultracold
plasmas were the so-called active space experiments, where the ionized
gas clouds were released into the vacuum from spacecraft and after a sharp
expansion could also evolve into the ultracold state~\cite{Dumin00a,Dumin01}.
Unfortunately, the diagnostic facilities in space were very limited as
compared to the laboratory installations.

If energy of the absorbed photon is approximately equal to the ionization
threshold of the atom, then the released photoelectron will possess zero
total energy at infinity (where its velocity tends to zero).
Therefore, when the electron is separated from the original atom,
\textit{e.g.}, by half the characteristic interionic distance~$ l/2 $
(in other words, it becomes to be governed by the ``collective'' plasma field),
its kinetic and potential energies should be equal to
\begin{equation}
k (l/2) = | u (l/2) | = 2\,e^2 \! / l \, .
\label{eq:electron_energies}
\end{equation}
In other words,
\begin{equation}
{\Gamma}_e \approx | u (l/2) | \, / \, k (l/2) = 1 \, ,
\label{eq:Gamma_estimate}
\end{equation}
\textit{i.e.}, the coupling parameter can be about unity but not
substantially larger than this value.
(In fact, due to additional heating, \textit{e.g.}, by the three-body
recombination, the experimentally-achieved temperature turns out to be a few
times greater.
So, the coupling parameter~$ {\Gamma}_e $ becomes only a fraction of unity:
for example, the recently reported values~\cite{Chen17} were up to~0.35.)

A natural question arises: Is it possible to overcome the
limit~(\ref{eq:Gamma_estimate})?
One of the most evident ideas is to irradiate the atoms by photons with
energy slightly below the ionization threshold.
In such a case, the atoms will not be ionized directly but transferred firstly
into the strongly-excited (Rydberg) states.
Next, these ``inflated'' atoms will be ionized spontaneously due to the
interparticle interactions.
In fact, the effect of spontaneous ionization of the ultracold Rydberg gas
and its transformation into the plasma was discovered long time
ago~\cite{Vitrant82}, and the same phenomenon was observed already in
the first experiments with the magneto--optical traps~\cite{Robinson00,Li05}.

\begin{figure}
\includegraphics[width=0.68\columnwidth]{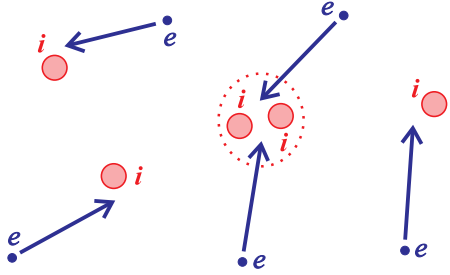}
\caption{\label{fig:Sketch_scattering}
Sketch of the simultaneous scattering of two electrons~($ e $) at
the ionic cluster, namely, two closely-located ions~($ i $),
marked by the dotted ring.}
\end{figure}

At first sight, it can be naturally expected that if the initial energy
introduced into the system becomes less, then the resulting electron
temperature should be substantially reduced and, therefore, considerable
values of the Coulomb coupling parameter, $ {\Gamma}_e\,{\gg}\,1 $, can be
achieved.
However, a closer analysis reveals the dangerous pitfall:
Namely, the ions produced by the spontaneous avalanche ionization will have
a strongly nonuniform (clusterized) distribution in space.
Then, as illustrated in Fig.~\ref{fig:Sketch_scattering}, such compact ionic
clusters (composed, for example, of two particles) will simultaneously
attract a few electrons and, thereby, stimulate the inelastic multi-particle
scattering, followed by the efficient redistribution of energy between
the particles.
(As is known, a bi-particle scattering of an electron by the ion cannot result
in a noticeable redistribution of energy: it is equivalent to the reflection
of a light particle from a ``wall'', which can change only direction of
the momentum of the scattered particle but not its absolute value and,
therefore, the kinetic energy.)
So, if multi-particle processes are allowed, the electron temperature
can increase substantially.

Yet another factor potentially leading to the increased temperature is that
the clusterized ions contribute more to the total value of the potential
energy appearing in the virial theorem~\cite{Landau76} for the system of
charged particles.
As a result, at the earliest stage of establishment of the quasi-equilibrium
energy distribution between the interacting particles (at the time scale about
the inverse plasma frequency), the virial value of the kinetic energy should
also be larger.
We shall discuss the relative contributions of both the above-mentioned
effects in more detail in the subsequent sections of the paper.

Therefore, it is unclear in advance if the reduced value of energy introduced
initially into the system will lead to the less electron temperature,
or it will be quickly compensated by the more violent subsequent relaxation.
It is the aim of the present work to answer this question.
Let us emphasize that we shall not try to give a self-consistent description
of the entire process of the spontaneous ionization.
Instead, we shall start from the predefined distributions of ions in space,
which are assumed to be formed due to the avalanche ionization, and then
study in much detail the subsequent relaxation of the electron velocities
in the corresponding ionic configurations.

\begin{figure}
\includegraphics[width=0.65\columnwidth]{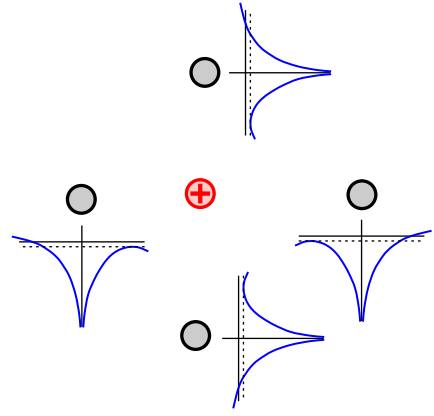}
\caption{\label{fig:Reduced_threshold}
Shift (asymmetry) of the potential curves of atoms in the vicinity of
a pre-formed ion, facilitating their subsequent ionization.}
\end{figure}

Let us briefly mention that, apart from the collisional avalanche ionization,
there might be yet another conceivable mechanism of the cluster formation
in the case when ions are produced by the near-threshold narrow-band
laser irradiation.
Namely, the first formed ion shifts the potential curves of nearby atoms,
thereby facilitating production of additional ions in the same place;
see Fig.~\ref{fig:Reduced_threshold}.

\begin{figure}
\includegraphics[width=0.8\columnwidth]{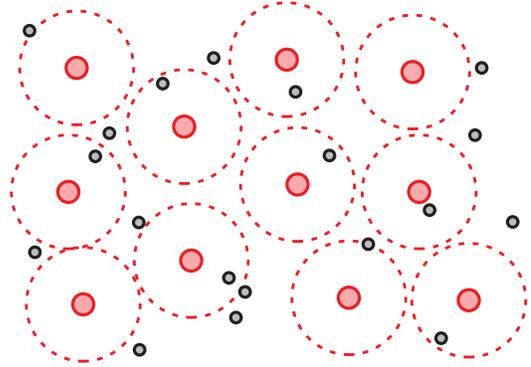}
\caption{\label{fig:Blockaded_gas}
Arrangement of Rydberg atoms (large red circles) in the blockade regime.
Dashed rings are the spots of the Rydberg blockade, and small black
circles are the unexcited (background) atoms.}
\end{figure}

One more problem to be addressed in the present paper is treating the case
of suppressed clusterization, which can be achieved by the two-step plasma
formation with the so-called effect of Rydberg blockade at the first
stage, as was suggested in Ref.~\onlinecite{Robert13}.
The Rydberg blockade is the phenomenon when one Rydberg atom, excited by
a narrow-band laser irradiation, shifts the energy levels of the nearby atoms
due to its huge electric dipole moment.
As a result, these levels turn out to be beyond the energy band of
the irradiation and cannot be excited.
So, a subsequent formation of the Rydberg atoms in the neighborhood of
the previously excited atom becomes
impossible~\cite{Lukin01,Tong04,Singer04}.
In principle, if the disturbance is sufficiently strong, the neighboring
energy levels can enter the excitation band of irradiation and, thereby,
to restore the possibility of excitation to the Rydberg
states~\cite{Keating13,Dumin14,Derevianko15}.
Moreover, in the case of plasmas, the shift of energy levels can be produced
also by the much stronger electric fields of ions.
As a result, these levels will either leave or enter the excitation band,
thereby leading to the phenomena of Coulomb blockade or anti-blockade,
respectively~\cite{Bounds19}.

Therefore, as illustrated in Fig.~\ref{fig:Blockaded_gas}, the blockaded
Rydberg atoms form a quasi-regular lattice: they are placed approximately at
the same distance from each other, so that the possibility of close
localization to each other is excluded.
If such atoms are subsequently irradiated by an ionizing pulse, then
the resulting ions will also possess the quasi-regular arrangement,
\textit{i.e.}, their clusterization will be efficiently suppressed.
(The atoms that were not initially excited because of the Rydberg blockade
remain unionized and do not participate in the plasma processes.)

The experimental attempt to produce plasmas by the two-step process with
Rydberg blockade was undertaken, \textit{e.g.}, in
Ref.~\onlinecite{Robert13}.
Unfortunately, because of the limited diagnostic capabilities,
it remained unclear if the electron temperature was really reduced or,
in other words, if this is a reasonable approach to the creation of
the strongly-coupled ultracold plasmas.
So, to answer this question, we shall numerically simulate below both
the cases of enhanced and suppressed ionic clusterization.

Besides, the quasi-regular arrangement of ions can be obtained also in
the optical lattices, formed by the counter-propagating laser beams.
This case was considered in the previous literature mostly in the context of
temporal behavior of the ionic coupling parameter~$ {\Gamma}_i $;
see, for example, Ref.~\onlinecite{Murphy16} and references therein.
However, it might be interesting to consider also dynamics of electrons in
such kind of the ionic background.

At last, a rather sophisticated manipulation with ionic distribution
functions can be performed due to specific features of Penning ionization
in the \textit{molecular} Rydberg gases~\cite{Sadeghi14}.

\section{Numerical Simulations}
\label{sec:Simulation}

\subsection{Formulation of the Model}
\label{sec:Model}

\begin{figure}
\includegraphics[width=0.98\columnwidth]{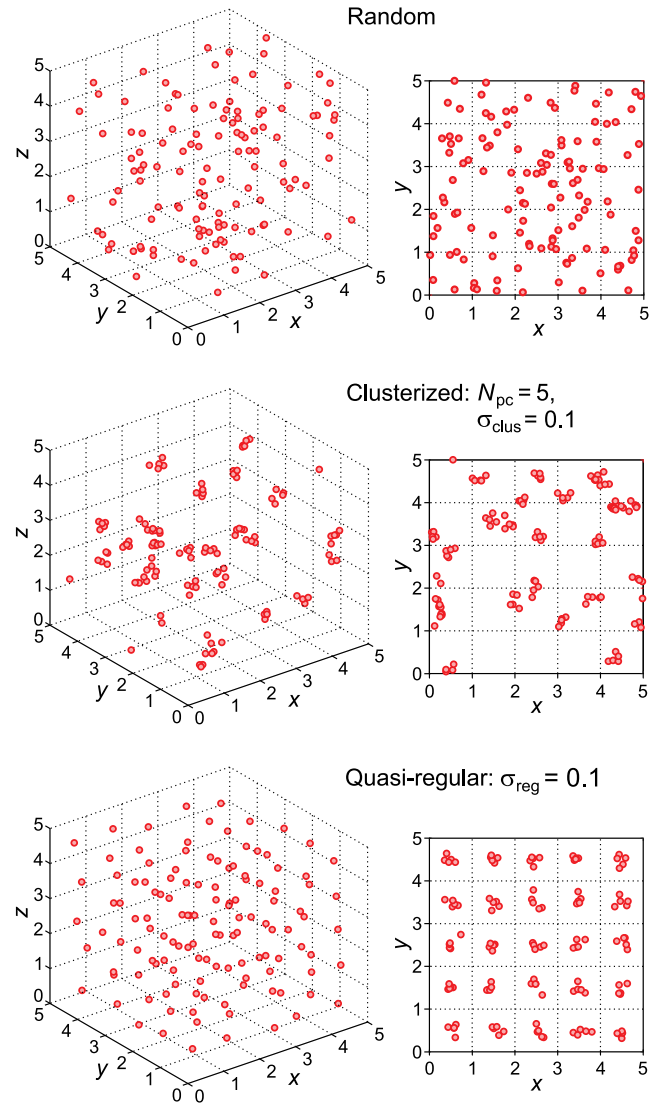}
\caption{\label{fig:Ionic_distribution}
Examples of the ionic arrangement (3D views on the left-hand side, and
$ xy $-projections on the right-hand side) for the statistically-uniform
random distribution (top row), enhanced clusterization with $ N_{\rm pc} = 5 $
and $ {\sigma}_{\rm clus} = 0.1 $ (middle row), and suppressed clusterization
with $ {\sigma}_{\rm reg} = 0.1 $ (bottom row).
All distances are measured in the units of~$ l $.}
\end{figure}

The main idea of subsequent simulations is to consider a relaxation of
the electron velocities against the background of immobile ions with various
kinds of their arrangement.
Namely, the following cases will be tested:
\begin{enumerate}
\item
Statistically-uniform distribution of ions, where some kind of clusterization
is possible only due to occasional coincidence of the ionic coordinates;
see Fig.~\ref{fig:Ionic_distribution} (top row).
As an example, we use here a cubic box composed of 5~cells of length~$ l $
in each direction, where each ``unitary'' cell (denoted by dotted lines)
contains \textit{on average} exactly one particle.
\item
Enhanced clusterization, where each cluster contains $ N_{\rm pc} $ ions
(the abbreviation ``pc'' implies ``particles per cluster'').
To get such an arrangement, we firstly take the centers of the clusters
to be distributed statistically uniform in space, and then positions of
the individual ions are generated according to the normal (Gaussian) law
with the root-mean-square (r.m.s.) deviation~$ {\sigma}_{\rm clus} $
with respect to the cluster centers;
see Fig.~\ref{fig:Ionic_distribution} (middle row).
\item
Suppressed clusterization, where ions were initially located in the nodes
of a perfectly regular cubic lattice with the cell size~$ l $ and then
randomly shifted from these positions (randomized) by the normal law
with r.m.s.\ deviation~$ {\sigma}_{\rm reg} $;
see Fig.~\ref{fig:Ionic_distribution} (bottom row).
\end{enumerate}
The grid of the dotted lines in these figures, denoting the unitary cells,
does not have a rigorous physical meaning in the cases of random and
clusterized distributions.
Nevertheless, this grid is very convenient for the visualization of the
average density: each cell corresponds on the average just to one particle.

In the numerical simulations, we used various values of
the parameters~$ N_{\rm pc} $, $ {\sigma}_{\rm clus} $, and
$ {\sigma}_{\rm reg} $, which will be specified below.
Besides, for each particular set of these parameters, the simulations were
performed for the sufficiently large number of initial conditions
(usually, 15) to make the results statistically significant.

Let us mention that the pictures of ionic arrangement are clearly distinct
from each other only at the relatively small values of r.m.s.\
deviations~$ {\sigma}_{\rm clus} $ and $ {\sigma}_{\rm reg} $, \textit{e.g.},
0.1 (in the units of the interparticle distance~$ l $), as in
Fig.~\ref{fig:Ionic_distribution}.
On the other hand, when the r.m.s.\ deviations are on the order of unity,
both the clusterized and quasi-regular distributions become quite similar
to the random one.
Moreover, as will be seen from the results of subsequent numerical simulations,
the electron dynamics in such cases will also be almost the same.

\begin{figure*}
\includegraphics[width=0.9\textwidth]{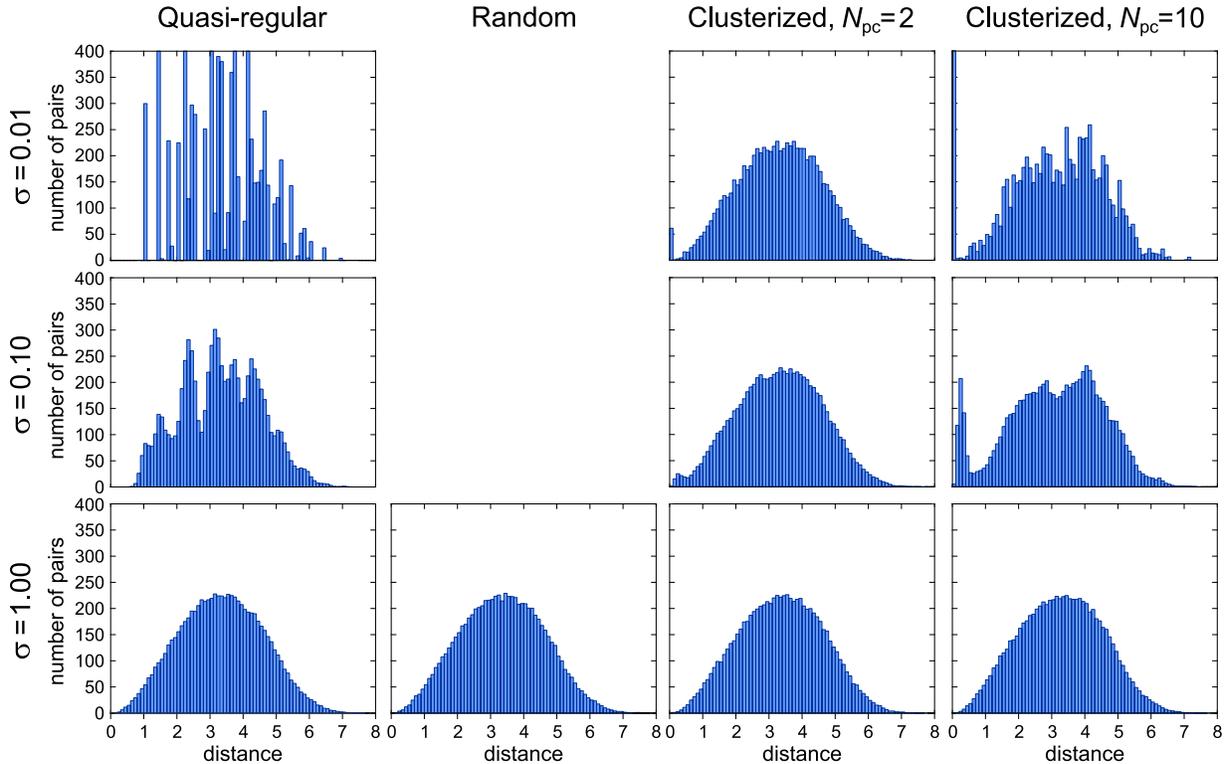}
\caption{\label{fig:Interparticle_hist}
Histograms of the interparticle separation for various kinds of
ionic arrangement.}
\end{figure*}

At last, it is interesting to look at the histograms of interparticle
separation (\textit{i.e.}, up to normalization, the pair correlation
functions), which are presented in Fig.~\ref{fig:Interparticle_hist}.
As should be expected, these histograms are strikingly different from
each other at small values of the r.m.s.\ deviation (especially, at
$ \sigma\,{=}\,0.01) $.
On the other hand, the histograms become very similar at the large value
of~$ \sigma $ (namely, $ \sigma\,{=}\,1.0 $): in this case, all particles
are randomly shifted from their original positions by the distances comparable
to the average interparticle separation.
As a result, all distributions are quasi-random, and their histograms take
the Gaussian-like shape.

In the case of quasi-regular distributions with small~$ {\sigma}_{\rm reg} $,
the histograms are very spiky, since the quasi-regular arrangement of particles
assumes a large set of the preferable distances between them, exactly as in
a crystalline structure.
Of course, when $ {\sigma}_{\rm reg} $ increases (\textit{i.e.},
the distribution is randomized), these spikes gradually disappear.
The most surprising fact is that, even when the entire distribution is very
spiky, this actually does not affect the relaxation of electron velocities,
as will be seen from the results of numerical simulations below.

On the other hand, a specific feature of the clusterized distributions is
a narrow bump (local maximum) at very small distances, on the order of
the typical scatter of particles inside the cluster~$ {\sigma}_{\rm clus} $.
While such a bump often looks like a minor disturbance of the histogram,
as will be seen from the subsequent simulations, it changes dramatically the
efficiency of relaxation of the electron velocities due to the effect of
multi-particle scattering, illustrated in Fig.~\ref{fig:Sketch_scattering}.

So, to simulate dynamics of the electrons against the above-mentioned ionic
backgrounds, we performed a straightforward numerical integration of
their equations of motion:
\begin{equation}
\frac{d^2}{dt^2} \, {\bf r}_i =
  - \sum_{j} e^2 \frac{ {\bf r}_i - {\bf R}_j }{ | {\bf r}_i - {\bf R}_j |^3 }
  + \sum_{k \neq i} e^2
    \frac{ {\bf r}_i - {\bf r}_k }{ | {\bf r}_i - {\bf r}_k |^3 } \, ,
\label{eq:Electron_motion}
\end{equation}
where
$ {\bf r}_i $ and $ {\bf R}_i $ are the coordinates of electrons and ions
respectively, and
$ e $~is the elementary electric charge.
Since we are interested only in the sufficiently short time scales,
the ionic motion was ignored (\textit{i.e.}, the ions were assumed to be
at rest at the time interval of the simulation).

While the initial arrangement of ions was outlined above, the initial
coordinates of electrons were always specified by a uniform statistical
distribution; and their velocities, by the normal (Gaussian) distribution
with r.m.s.\ deviation~$ {\sigma}_{\! v} $.
In the particular simulations presented below, we used
$ {\sigma}_{\! v} = 0.3 $, which means that the initial kinetic energy of
electrons was an order of magnitude less than their potential (Coulomb)
energy.
So, the plasma at $ t\,{=}\,0 $ was assumed to be ``overcooled''.

Strictly speaking, the random initial positions for the electrons
are somewhat artificial.
In the case of instantaneous photoionization, the electron positions should
be initially strongly correlated with the positions of ions; and just this
case was simulated in detail in the work~\cite{Niffenegger11}.
On the other hand, if the ionization process takes some time, the
released electrons will be mixed in space between the ions.
So, one can expect that their positions might be reasonably described
by the uniform random distribution.
As follows from the comparison of our subsequent results with the
above-cited paper, a temporal dependence of the kinetic energy turns
out to be qualitatively the same, apart from the very early time
interval (about the inverse plasma frequency).

For simplicity, the perfectly reflective conditions were imposed at
the boundaries of the simulation box.
As an alternative, we tried to use also the periodic boundary conditions.
Unfortunately, they required a much more computational time to get the
convergent results, which was unacceptable in the present study;
for more details, see Appendix~\ref{sec:Bound_cond}.

Of course, for the simulation with a relatively small number of particles
within a box with reflective boundaries to correspond to the physical
reality, it is necessary that the Debye screening length~$ r_{\rm D} $
be much smaller than the box size~$ L $.
Since $ r_{\rm D} / l = \sqrt{ K / 6 \pi U} $ (where
$ K = (3/2)\,k_{\rm B} T_e $~is electron kinetic energy and
$ U \approx e^2 / l $~ is the potential one) and $ L = 5\,l $,
one can easily find that at the initial instant of time, when we
specified $ K \approx 0.1\,U $, the above-mentioned ratio was
$ r_{\rm D} / L \approx 0.015 $.
Next, when the electrons are heated in the course of their subsequent
dynamics, the temperature can increase by 300 times, up to
$ K \approx 30\,U $ (see Table~\ref{tab:Resulting_temperature} and
Fig.~\ref{fig:Resulting_temperature} below).
Then, $ r_{\rm D} / L \approx 0.26 \ll 1 $, \textit{i.e.}, the required
inequality is still satisfied.

The set of equations~(\ref{eq:Electron_motion}) was integrated by
the numerical algorithm with the adaptive stepsize control, based on
the combination of Runge--Kutta methods of the 4th and 5th order
(subroutines \texttt{odeint}, \texttt{rkck}, and \texttt{rkqs} from
the book~\cite{Press92}).
This enabled us to deal with the ``original'' (singular) Coulomb potentials,
without any artificial cut-off or ``softening'' at the very small
distances, \textit{e.g.}\ Refs.~\onlinecite{Niffenegger11,Tiwari17}.
Thereby, all artifacts caused by the distorted potentials were completely
excluded.
In this sense, our simulation follows the recent
works~\cite{Venkatesh20,Bobrov20,Bronin21} as well as our earlier
study~\cite{Dumin11}.
However, despite dealing with the singular functions, the adaptive stepsize
control provided us the accuracy of integration (\textit{e.g.}, estimated
by the conservation of energy) at the same level as in the majority of other
simulations of ultracold plasmas with softened or truncated potentials.
Namely, the error was usually about~0.1{\%} and very rarely increased up
to~5{\%}.

To check the accuracy of simulations, we tried to perform calculations
with various number of particles in the simulation cell.
The major part of our results for the variety of initial parameters
presented below were obtained for the case of only $ N_{\rm tot} = 125 $
electrons and 125~ions, \textit{i.e.}, exactly for the situation
illustrated in Fig.~\ref{fig:Ionic_distribution}.
Increasing the number of particles, \textit{e.g.}, up to $ N_{\rm tot} = 1000 $
requires much more computational time (which scales approximately as
$ N_{\rm tot}^2 $), but the resulting average curves remain almost the same;
for more details, see Appendix~\ref{sec:Num_part}.
This is in accordance with the earlier study~\cite{Niffenegger11}, where
a hundred of particles of each kind were found to be sufficient to simulate
the electron temperature with a reasonable accuracy.

For the sake of brevity, we shall use below the dimensionless quantities,
normalized to the following basic units:
the unit of length is the size of the ``unitary'' cell (or mean distance
between the ions)~$ l $;
the unit of time is inversely proportional to the square root of the plasma
frequency,
$ \tau = \sqrt{m l^3 / e^2} = \sqrt{4 \pi} / {\omega}_{\rm pl} $;
and the unit of energy is the characteristic Coulomb energy,
$ U = e^2 / l $.
The corresponding normalized quantities will be denoted by hats.

We shall assume below that the electron kinetic energy per particle~$ K $
(with coefficient 2/3) is a direct measure of the electron
temperature~$ T_e $.
It was found in Ref.~\onlinecite{Niffenegger11} that such definition
of~$ T_e $ reasonably coincides with a more sophisticated derivation of~$ T_e $
by fitting the simulated velocity distributions to the Maxwellian ones,
provided that the electrons located sufficiently close to ions
(within the ``exclusion sphere'') are not taken into account in
calculation of the kinetic energy.
In fact, as was discussed in our earlier work~\cite{Dumin01},
the straightforward definition $ T_e = (2/3)\,K / k_{\rm B} $ should work
rather well even for the electrons strongly interacting with ions.
Besides, the introduction of the exclusion spheres is not a self-consistent
procedure: a small distance of an electron from the nearby ion at some
instant of time cannot be a criterion of its capture by this ion.
This is the reason why we take into account all electrons in calculation
of the kinetic energy.

\subsection{Results of the Numerical Calculations}
\label{sec:Results}

\begin{figure}
\includegraphics[width=0.9\columnwidth]{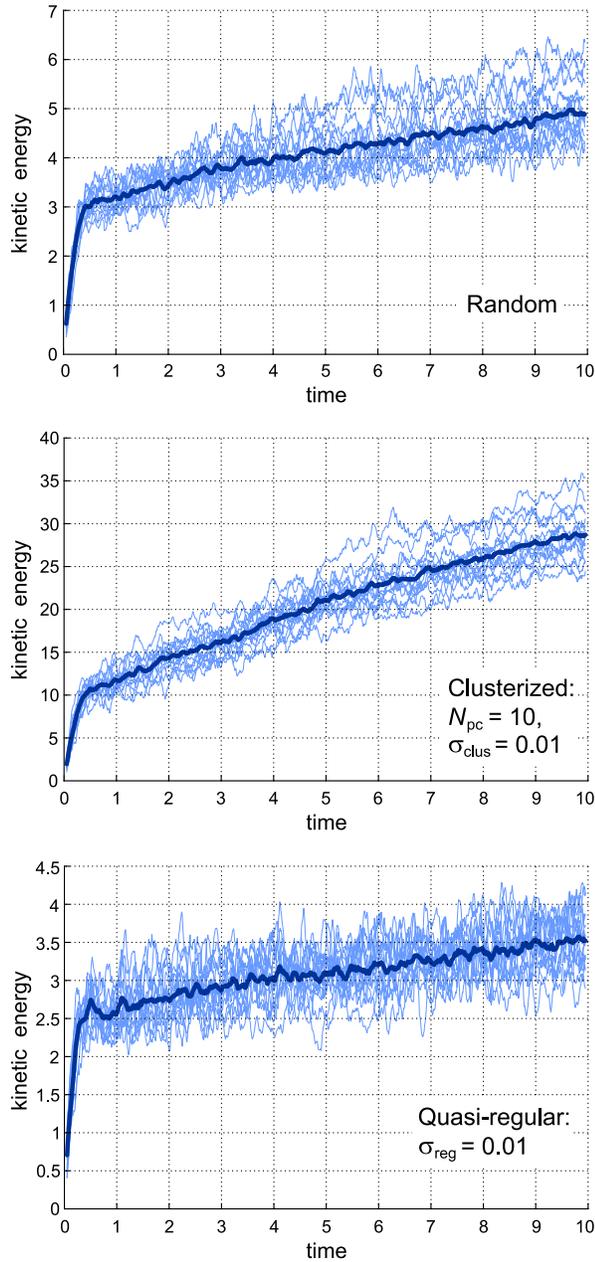}
\caption{\label{fig:Energy_Time}
Examples of the temporal behavior of the electron kinetic energy
in the case of the purely random ionic distribution (top panel),
enhanced clusterization of ions (middle panel), and
the suppressed clusterization (bottom panel).
Thin blue curves show the individual simulations with different
initial conditions for $ N_{\rm tot} = 125 $ particles of each kind,
and the thick curve is their average over 15~realizations.}
\end{figure}

Examples of the computed temporal behavior of the electron kinetic energy for
various kinds of ionic arrangement are shown in Fig.~\ref{fig:Energy_Time}.
To avoid cluttering the figure with a lot of sharp peaks, caused by
the close passages of electrons near the ions, the data were smoothed out
over the running window of width $ {\Delta}\hat{t} = 0.1 $.

In the case of purely random ionic distribution (top panel), this energy
quickly jumps approximately up to the virial value at the time scale
$ \hat{t} \sim 0.5 $, which is in agreement with the pioneering
work~\cite{Kuzmin02}, as well as with our earlier simulations~\cite{Dumin11}.
At longer times, the energy continues to increase but much more
slowly, due to the heat release by the three- (or multiple-)body recombination,
when some electrons become permanently captured by ions.

Next, in the case of considerable clusterization (\textit{e.g.},
$ N_{\rm pc} = 10, {\sigma}_{\rm clus} = 0.01 $, middle panel),
both the initial jump of energy is a few times greater, and its subsequent
increase is much more pronounced.
Both these features are not surprising:
Really, since the potential energy of compact clusters is larger than in
the random distribution, the kinetic energy established after
the ``virialization'' (\textit{i.e.}, after the first stage of relaxation)
should be also larger.
Besides, since the above-mentioned ionic clusters act as the multiply-charged
centers of attraction for a few electrons
(see Fig.~\ref{fig:Sketch_scattering}), the efficiency of the three- and
multiple-body recombination should also increase, leading to more pronounced
heating at the second stage.
Naturally, the total increase of the electron temperature becomes larger
for the clusters with larger number of ions~$ N_{\rm pc} $
and the smaller size~$ {\sigma}_{\rm clus} $, as will be seen below in
Fig.~\ref{fig:Energy_Time_groups}.

At last, behavior of the electron kinetic energy for the suppressed
clusterization of ions (bottom panel) looks quite similar to the case of
the purely random distribution.
In fact, the resulting values of the electron temperature are a bit smaller
than for the random distribution; but this difference is not so significant.
A very nontrivial finding of our simulations is that the above-mentioned
similarity with random distribution persists even at the very small values
of~$ {\sigma}_{\rm reg} $, corresponding to the almost regular (weakly
distorted) ionic lattices.

\begin{figure}
\includegraphics[width=0.9\columnwidth]{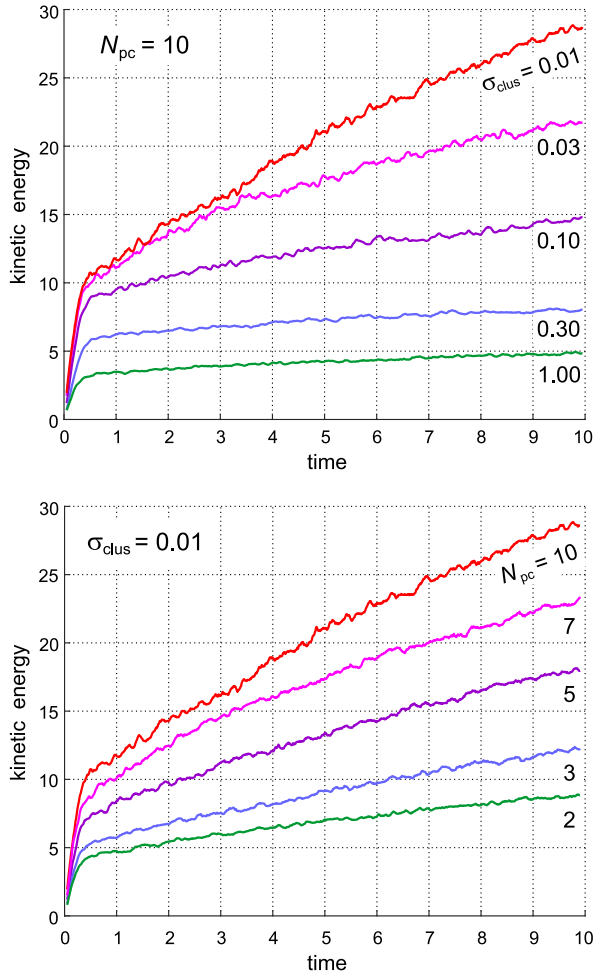}
\caption{\label{fig:Energy_Time_groups}
The average (over 15~realizations) temporal behavior of the electron
kinetic energy at the fixed~$ N_{\rm pc}{=}10 $ and
various~$ {\sigma}_{\rm clus} $ (top panel) and, \textit{vice versa},
the fixed~$ {\sigma}_{\rm clus}{=}0.01 $ and various~$ N_{\rm pc} $
(bottom panel).}
\end{figure}

To study the case of enhanced clusterization in more detail,
it is insightful to plot simultaneously the curves of kinetic energy at
the fixed number of particles per cluster~$ N_{\rm pc} $ (\textit{e.g.}, 10)
and various~$ {\sigma}_{\rm clus} $ (namely, 1, 0.3, 0.1, 0.03, and 0.01),
on the one hand, and the same curves at the fixed r.m.s.\ deviation of
particles from the cluster center~$ {\sigma}_{\rm clus} $ (\textit{e.g.}, 0.01)
and various~$ N_{\rm pc} $ (namely, 2, 3, 5, 7, and 10), on the other hand.
As is seen in the top panel of Fig.~\ref{fig:Energy_Time_groups}, when
$ N_{\rm pc} $ is constant and $ {\sigma}_{\rm clus} $ decreases from~1
to~0.03 (\textit{i.e.}, the clusters become more compact), the initial jump
of temperature (at the time interval from 0 to approximately 0.3) considerably
increases, but the subsequent heating (at $ \hat{t} \gtrsim 0.3 $) changes not
so appreciably.
However, when $ {\sigma}_{\rm clus} $ decreases further to~0.01, the initial
jump is ``saturated'', but the subsequent heating begins to operate
more efficiently.

On the other hand, as is seen in the bottom panel of
Fig.~\ref{fig:Energy_Time_groups}, when $ {\sigma}_{\rm clus} $ is constant
while the number of particles per cluster~$ N_{\rm pc} $ increases from~2
to~10, then both the initial jump of the temperature and the subsequent slope
of the curves increase simultaneously.

\begin{figure}
\includegraphics[width=0.98\columnwidth]{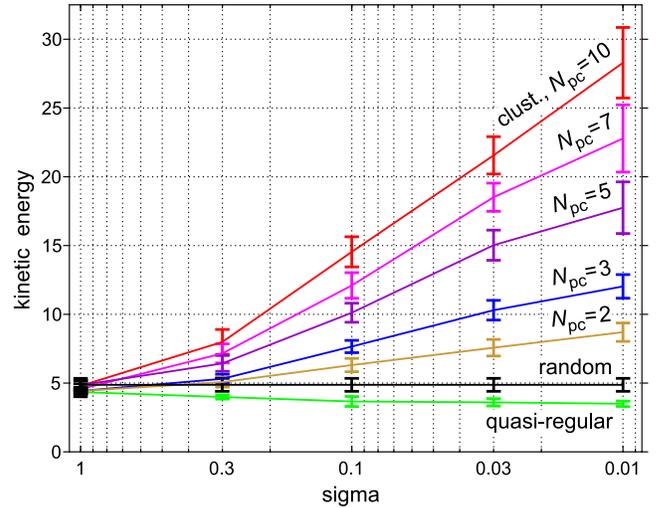}
\caption{\label{fig:Resulting_temperature}
Resulting values of the electron kinetic energy (in dimensionless units),
formed at the interval $ \hat{t} \in [9, 10] $, for various types of
the background ionic distribution.
Here, $ \sigma $ implies either $ {\sigma}_{\rm clus} $ or
$ {\sigma}_{\rm reg} $, depending on the particular context;
and their values are plotted in the decreasing order.
Vertical bars denote the r.m.s.\ deviations over the various versions
of initial conditions~$ \hat{\sigma}_{\rm ver} $.}
\end{figure}

Finally, the most interesting item for the possibility of creation of
the strongly-coupled plasmas are the resulting values of the electron
kinetic energy~$ \langle \hat{K} \rangle $ after a sufficiently long period
of evolution.
So, it is insightful to plot these values as function of $ N_{\rm pc} $
and $ {\sigma}_{\rm clus} $.
Figure~\ref{fig:Resulting_temperature} represents the corresponding quantities
averaged over the time interval $ \hat{t} \in [9, 10] $ (which was taken
somewhat arbitrary) and 15~different versions of initial conditions.
The same data are listed in more detail in
Table~\ref{tab:Resulting_temperature}.
Strictly speaking, parameter~$ \sigma $ is irrelevant to
the purely random distribution.
Nevertheless, the corresponding data are formally placed in the column with
$ {\sigma}_{\rm reg/clus} = 1 $, because in this case both the clusterized and
quasi-regular distributions become very similar to the random one.
To characterize ``stability'' of the average values, both the r.m.s.\
deviations with respect to time (over the above-specified
interval)~$ \hat{\sigma}_t $ and with respect to different versions of
the initial conditions~$ \hat{\sigma}_{\rm ver} $ are presented there.
As is seen, the values of~$ \hat{\sigma}_t $ are usually (but not always)
somewhat larger than~$ \hat{\sigma}_{\rm ver} $.
To avoid cluttering Fig.~\ref{fig:Resulting_temperature},
only~$ \hat{\sigma}_{\rm ver} $ are plotted by the vertical bars.

\begin{table*}
\caption{\label{tab:Resulting_temperature}
Values of the electron kinetic energy
$ \langle \hat{K} \rangle \pm \hat{\sigma}_t \pm \hat{\sigma}_{\rm ver} $
achieved in the interval $ \hat{t} \in [9, 10] $, where
$ \hat{\sigma}_t $~is the r.m.s.\ deviation with respect to time,
and $ \hat{\sigma}_{\rm ver} $~with respect to various versions of
initial conditions}
\begin{ruledtabular}
\begin{tabular}{cllrrrrrc}
\quad &
\multicolumn{2}{c}{$ {\sigma}_{\rm reg/clus} $} &
\multicolumn{1}{c}{0.01} &
\multicolumn{1}{c}{0.03} &
\multicolumn{1}{c}{0.10} &
\multicolumn{1}{c}{0.30} &
\multicolumn{1}{c}{1.00} &
\quad
\\[0.8ex]
\hline
\\[-1.4ex]
& \multicolumn{2}{l}{Random} &
  --------------------- & --------------------- &
  --------------------- & --------------------- &
  $ 4.87{\pm}0.72{\pm}0.57 $ &
\\
& \multicolumn{2}{l}{Quasi-regular} &
$ 3.50{\pm}0.61{\pm}0.29 $ &
$ 3.60{\pm}0.66{\pm}0.35 $ &
$ 3.66{\pm}0.63{\pm}0.46 $ &
$ 3.99{\pm}0.64{\pm}0.24 $ &
$ 4.35{\pm}0.60{\pm}0.43 $ &
\\
& Clusterized: & $ N_{\rm pc}\! = 2 $ &
$ 8.70{\pm}1.15{\pm}0.75 $ &
$ 7.57{\pm}1.03{\pm}0.69 $ &
$ 6.31{\pm}0.83{\pm}0.57 $ &
$ 5.07{\pm}0.68{\pm}0.44 $ &
$ 4.42{\pm}0.72{\pm}0.28 $ &
\\
&& $ N_{\rm pc}\! = 3 $ &
$ 12.03{\pm}1.83{\pm}0.94 $ &
$ 10.30{\pm}1.23{\pm}0.80 $ &
$  7.66{\pm}0.80{\pm}0.53 $ &
$  5.31{\pm}0.63{\pm}0.40 $ &
$  4.43{\pm}0.64{\pm}0.30 $ &
\\
&& $ N_{\rm pc}\! = 5 $ &
$ 17.75{\pm}2.46{\pm}1.97 $ &
$ 15.03{\pm}1.72{\pm}1.17 $ &
$ 10.12{\pm}1.04{\pm}0.77 $ &
$  6.43{\pm}0.83{\pm}0.65 $ &
$  4.85{\pm}0.71{\pm}0.44 $ &
\\
&& $ N_{\rm pc}\! = 7 $ &
$ 22.79{\pm}2.89{\pm}2.54 $ &
$ 18.51{\pm}1.86{\pm}1.10 $ &
$ 12.10{\pm}1.29{\pm}1.01 $ &
$  7.16{\pm}0.74{\pm}0.78 $ &
$  4.56{\pm}0.70{\pm}0.46 $ &
\\
&& $ N_{\rm pc}\! = 10 $ &
$ 28.29{\pm}3.52{\pm}2.64 $ &
$ 21.56{\pm}2.02{\pm}1.44 $ &
$ 14.54{\pm}1.35{\pm}1.18 $ &
$  7.99{\pm}0.69{\pm}0.98 $ &
$  4.80{\pm}0.63{\pm}0.46 $ &
\\
\end{tabular}
\end{ruledtabular}
\end{table*}

The quite large values of~$ \hat{\sigma}_t $ and
$ \hat{\sigma}_{\rm ver} $ were actually caused by the relatively small
number of the simulated particles.
If this number is increased (\textit{e.g.}, from $ N_{\rm tot} = 125 $ to
1000~particles of each kind), the r.m.s.\ deviations become much smaller,
but the average values almost do not change; for more details, see
Appendix~\ref{sec:Num_part}.

\section{Discussion and Conclusions}
\label{sec:Discussion}

As distinct from a number of previous theoretical studies of the avalanche
ionization, which were based on the kinetic rate equations (\textit{e.g.},
review~\cite{Aghigh20}), we performed a self-consistent \textit{ab initio}
modeling of the many-body effects in the plasmas with a nontrivial ion
arrangement.
(It is interesting to mention that already the pioneering work on the
avalanche ionization of Rydberg gas~\cite{Vitrant82} emphasized that
``it is thus possible that the two-body analysis is too naive''.)
Thereby, we arrived at the following conclusions:

\begin{enumerate}

\item
Studying both the cases of enhanced and suppressed clusterization, in general,
reveals the same features as in the previous works: if system of the charged
particles is taken initially in the state with very small kinetic energy
(\textit{e.g.}, by an order of magnitude less than the potential one), then
the kinetic energy begins to increase quickly.
This process proceeds in two stages:
Firstly, the electrons are sharply accelerated in the local electric fields
of the nearby ions and, thereby, increase their kinetic energy approximately
up to the virial value (\textit{i.e.}, about a half of the potential one).
This takes place on the time scale about the inverse plasma
frequency~$ {\omega}_{\rm pl} $ or, up to the numerical factor, the Keplerian
period (more exactly, $ \hat{t} \approx $ 0.3).
Next, the process of three- or multi-body recombination comes into play,
resulting in the subsequent gradual increase of the electron kinetic energy.

\item
As follows from Fig.~\ref{fig:Energy_Time_groups}, both stages of heating
proceed more intensively with increasing the number of particles (ions) per
cluster~$ N_{\rm pc} $ and decreasing the characteristic cluster
size~$ {\sigma}_{\rm clus} $.
Besides, according to the top panel of this figure, when $ N_{\rm pc} $ is
fixed, the increasing compactness of the clusters results initially in
more intense virialization (\textit{i.e.}, a sharper jump of the kinetic
energy at small~$ \hat{t} $)
and then to the more efficient heating due to the multi-particle
recombination, \textit{i.e.}, a steeper increase of the kinetic energy at
the second stage.
(To avoid misunderstanding, let us emphasize that---although we
do not take into account motion of the ions during the simulations---the
virial relations should be applied to the total Coulomb energy of all
charged particles, \textit{i.e.} the increased compactness of ionic
clusters should be of primary importance.)
Of course, the total increase of the kinetic energy at the sufficiently
large times (\textit{e.g.}, $ \hat{t} \in [9, 10] $) will depend
on~$ N_{\rm pc} $ and~$ {\sigma}_{\rm clus} $ in the same way;
see Fig.~\ref{fig:Resulting_temperature}.

\item
A very nontrivial finding of our simulations is that a suppressed
clusterization (modeled by the quasi-regular ionic distributions)
hardly affects the relaxation of the electron velocities.
As demonstrated by the lowest curve in Fig.~\ref{fig:Resulting_temperature},
the electron kinetic energy in the interval $ \hat{t} \in [9, 10] $ remains
almost the same when the r.m.s.\ shift of the ions from the regular lattice
positions~$ {\sigma}_{\rm reg} $ decreases by two orders of magnitude,
from 1 to 0.01; and it equals approximately the value for a purely random
distribution of ions.
Let us emphasize that this conclusion refers only to the electron
temperature, while ionic temperature at the longer time scale can exhibit
a much more nontrivial behavior, depending on the degree of
disorder~\cite{Pohl04,Pohl05,Donko09,Murphy15,Murphy16}.
Particularly, suppression of the disorder-induced heating of ions,
\textit{e.g.}, by the Rydberg blockade of the original gas can be really
efficient.

\item
Referring to the histograms of interparticle separation (or, which is the
same, the pair correlation functions) in Fig.~\ref{fig:Interparticle_hist},
we see that at $ {\sigma}_{\rm clus} = {\sigma}_{\rm reg} = 1 $ the plots
are approximately Gaussian.
This is not surprising because at $ {\sigma}_{\rm clus} = 1 $ the clusters
become very diffuse, and at $ {\sigma}_{\rm reg} = 1 $ the original regular
lattice becomes completely destroyed.
Therefore, these random distributions become statistically uniform in space,
and the resulting values of the electron kinetic energy, presented in
Fig.~\ref{fig:Resulting_temperature} and Table~\ref{tab:Resulting_temperature},
are almost the same.
Next, when for the clusterized distribution $ {\sigma}_{\rm clus} $ decreases,
the Gaussian shape of the plots is slightly distorted, and a very narrow peak
is formed near zero (\textit{i.e.}, at the very small interparticle
separation).
In fact, just this peak is of crucial importance for the electron heating,
because it strongly enhances both the efficiency of initial virialization
and the subsequent multi-body recombination.
On the other hand, when for the quasi-regular distribution
$ {\sigma}_{\rm reg} $ decreases from 1 to 0.01, the Gaussian shape of
the plots becomes completely distorted, and a lot of the narrow peaks are
formed at various interparticle separations.
However, none of these peaks is localized near zero and, as a result, they are
insignificant for relaxation of the electron velocities.

\item
The main conclusion following from our simulations is that the clusterization
of ions (\textit{e.g.}, in the course of avalanche ionization) is a very
serious obstacle to getting large values of the Coulomb coupling
parameter~$ {\Gamma}_e $.
Really, since all energies were normalized to the characteristic Coulomb
energy of interparticle interaction (\textit{i.e.},
$ \langle \hat{U} \rangle \sim 1 $), large values of the dimensionless
kinetic energy~$ \langle \hat{K} \rangle $ obtained in the simulations imply
that the coupling parameter $ {\Gamma}_e \approx U/K = \hat{U}/\hat{K} $
will be rather small as compared to unity.
Therefore, the smaller energy input into the Rydberg gas (as compared to
the direct photoionization) will lose any advantage after formation of
clusters in the spontaneously-ionized plasma.
(Yet another conceivable method to reduce the electron temperature in
ultracold plasmas is to add there the Rydberg atoms with binding energies
$ | E_{\rm b} | \lesssim (2-3)\,k_{\rm B}T_e $.
Then, their inelastic collisions with electrons will lead to a further
excitation of the atoms and cooling of the electrons.
Unfortunately, as follows both from the experiments and numerical
simulations~\cite{Vanhaecke05,Crockett18}, the overall efficiency of such
process is quite low---the electron temperature can be reduced by
no more than 20{--}30\,\%.)

\item
At last, if the two-step formation of ultracold plasma involves the Rydberg
blockade at the first stage, resulting in the quasi-regular arrangement
of ions~\cite{Robert13}, the electron temperature and the corresponding
Coulomb coupling parameter~$ {\Gamma}_e $ should be stabilized approximately
at the same level as for the purely random ionic distribution.
Therefore, one should not expect that this method can appreciably increase
the attainable values of the coupling parameter.

\end{enumerate}

\begin{acknowledgments}

YVD is grateful to S.~Whitlock for drawing his attention to the problem of
temperature evolution in the blockaded Rydberg plasmas, as well as to
J.-M.~Rost for emphasizing the importance of clusterization in ultracold
plasmas.
We are also grateful to A.A.~Bobrov, S.A.~Mayorov, U.~Saalmann, and
V.S.~Vorob'ev for fruitful discussions and valuable suggestions.

\end{acknowledgments}

\section*{Author Declarations}

\subsection*{Conflict of Interest}

The authors have no conflicts to disclose.

\subsection*{Author Contributions}

YVD suggested the theoretical concept and developed the corresponding
software; both authors performed the simulations, analyzed the results,
and prepared the manuscript.

\section*{Data Availability}

The data that support the findings of this study are available from the
corresponding author upon reasonable request.

\appendix

\section{Effect of the Boundary Conditions on the Results of Simulations}
\label{sec:Bound_cond}

\begin{figure}
\includegraphics[width=0.9\columnwidth]{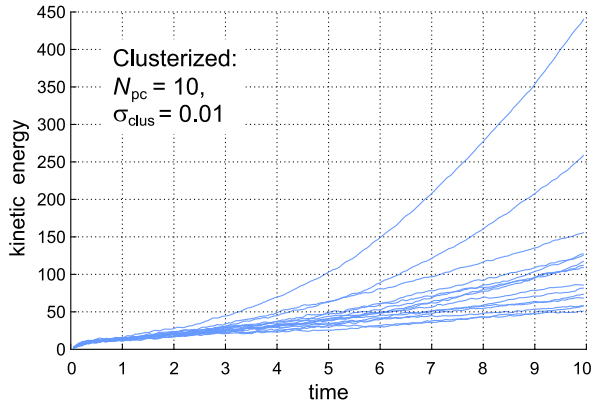}
\caption{\label{fig:Period_Bound}
Examples of the temporal behavior of the electron kinetic energy
in the case of periodic boundary conditions.}
\end{figure}

Since we performed all simulations with a relatively small number of
charged particles, it is of crucial importance to specify the adequate
boundary conditions.
At the first sight, the most reasonable choice would be the periodic
conditions, when a particle leaving the simulation box through some
boundary simultaneously enters this box from the same point at the
opposite boundary.
Unfortunately, our computations with periodic boundary conditions led
to the quite unsatisfactory results, whose example is presented in
Fig.~\ref{fig:Period_Bound}.
Namely, there was a huge scatter between the individual curves,
so that it was meaningless to draw the average curve as in
Fig.~\ref{fig:Energy_Time}.

\begin{figure*}
\includegraphics[width=0.85\textwidth]{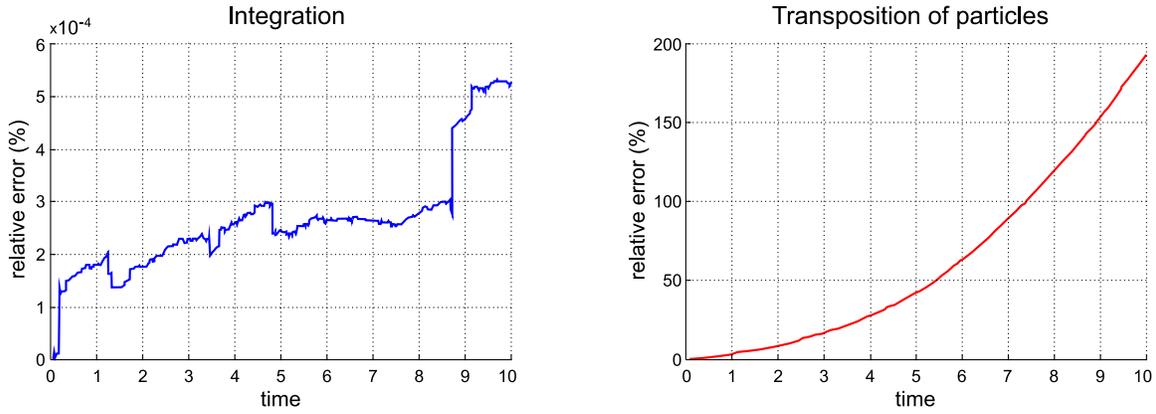}
\caption{\label{fig:Accum_Errors}
Example of accumulation of the errors due to the integration itself
(left panel) and due to the transposition of particles (right panel)
in one of the worst cases of initial conditions.}
\end{figure*}

A more careful analysis shows that the accumulation of errors
(estimated by conservation of the total energy) comes mostly from
the transposition of particles between the opposite boundaries,
while accuracy of the integration algorithm itself remains very good;
see Fig.~\ref{fig:Accum_Errors}.
In fact, the spread of curves in Fig.~\ref{fig:Period_Bound} is caused
by a population of the ``transient'' particles.
They are strongly accelerated at the very early stage of plasma
relaxation, when almost immobile electrons begin to fall onto the nearest
ions.
Subsequently, these ``transient'' electrons cross the boundaries of the
simulated volume many times, thereby accumulating the computational errors
due to the jumps of the Coulomb forces and energies during the
transpositions.

One possible remedy to exclude such errors is to increase the number of
``mirror'' boxes~$ N_{\rm m} $ in each direction from the ``basic'' box,
until the required accuracy of convergence of the Coulomb force and
energy is achieved.
This seems to be the most self-consistent method to take into account
the long-range character of the Coulomb interactions.
Particularly, just this approach was implemented in one of our earlier
works~\cite{Dumin11}.
Unfortunately, it is extremely time-consuming:
the total number of the mirror boxes to be included into the
calculations scales with~$ N_{\rm m} $ as $ (2 N_{\rm m} + 1)^3 $.
As follows from our previous experience~\cite{Dumin11}, the required
value of~$ N_{\rm m} $ at a particular step of integration changed
from 9 to 27, leading to the total number of mirror boxes to be processed
about~$ {10^4}-{10^5} $.
As a result, the entire simulation for a single set of initial conditions
took up to a few months of computational time.
Such computing requirements were evidently unacceptable in the present
work, because we aimed to perform simulations for a very large set of
initial conditions.
Therefore, we chose the reflective boundary conditions, where all
calculations (both integration of the equations of motion and the summation
of the Coulomb interactions) are performed within a single simulation box.

An alternative option might be to employ the ``wrap'' boundary
conditions~\cite{Niffenegger11}, where any particle interacts with other
particles within the cube of size~$ \pm L/2 $ centered at that particle.
It can be easily shown that such prescription ensures a smooth variation of
the Coulomb energy when the above-mentioned cube is redefined in the course
of interparticle displacements.
However, the corresponding Coulomb force experiences an abrupt unphysical
change (namely, its component perpendicular to the cube boundary suddenly
changes its sign for the particular pair of particles).
Moreover, the effect of such jumps cannot be represented by the error
estimated from the conservation of energy.

In summary, none of the boundary conditions for simulation of the infinite
system is perfect: each of them has its own advantages and disadvantages.
Anyway, we preferred to use here the reflective boundary conditions because
the abrupt changes in velocity, caused by the reflection, have a much more
physical meaning than the abrupt changes in the Coulomb forces.

\section{Effect of the Number of Particles on the Results of Simulations}
\label{sec:Num_part}

\begin{figure}
\includegraphics[width=0.9\columnwidth]{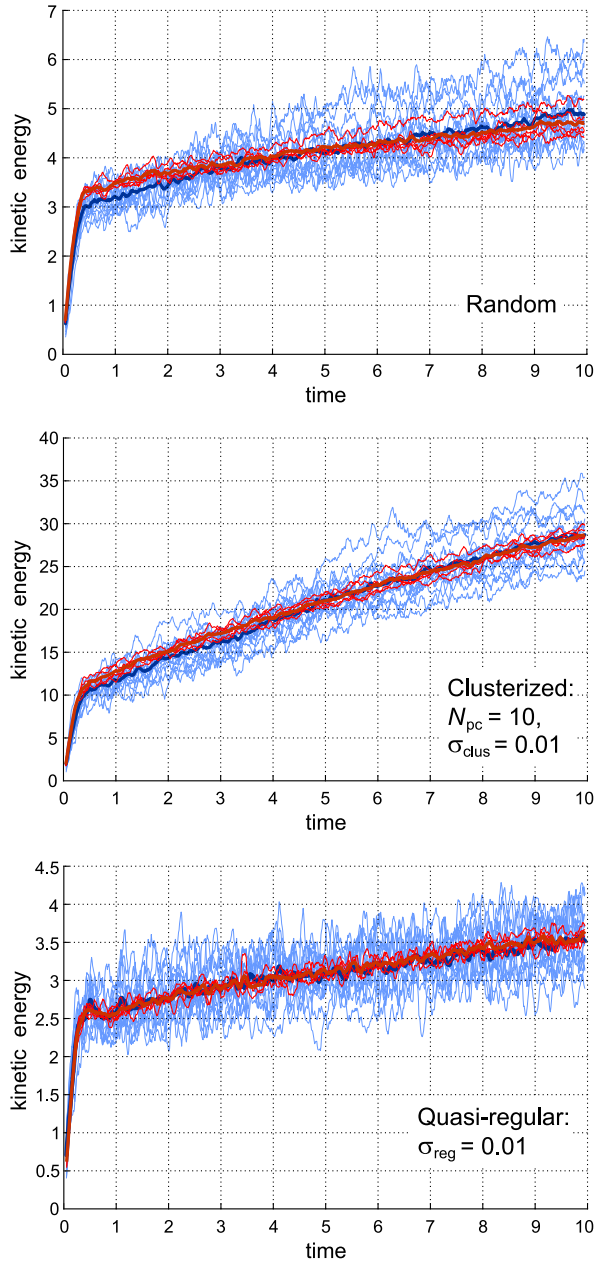}
\caption{\label{fig:Depend_N}
The same as in Fig.~\ref{fig:Energy_Time}, but with the results of
5~simulations for $ N_{\rm tot} = 1000 $ particles (red curves)
superimposed onto the results of 15~simulations for
$ N_{\rm tot} = 125 $ particles (blue curves).
The thin and thick curves show the individual simulations and
the average results, respectively.}
\end{figure}

Yet another important issue for the validity of simulations is to check that
the results depend only weakly on the total number of particles.
With this aim in view, we increased the size of the simulated box from
$ L = 5\,l $ to $ L = 10\,l $ and, respectively, the number of particles
from $ N_{\rm tot} = 125 $ to 1000, \textit{i.e.}, almost by an order of
magnitude.
The computations were repeated for 5~versions of initial conditions in each of
the three extreme cases: a completely random distribution of ions, the most
clusterized distribution (the number of ions per cluster $ N_{\rm pc} = 10 $
and their r.m.s.\ deviation within the cluster $ {\sigma}_{\rm clus} = 0.01 $),
and the almost regular distribution (the r.m.s.\ deviation with respect to
the ideal lattice $ {\sigma}_{\rm reg} = 0.01 $).
The corresponding results, combined with the previous simulations, are presented
in Fig.~\ref{fig:Depend_N}.
As is seen, the curves for $ N_{\rm tot} = 1000 $, drawn in red, possess
a substantially less r.m.s.\ deviation than the blue curves for
$ N_{\rm tot} = 125 $.
However, the average (thick) curves coincide with each other surprisingly well.
In fact, the blue curve is often invisible at all, because it is completely
covered by the red one.

The only noticeable difference between the average curves can be seen in
the top panel, referring to the completely random distribution of ions:
The red curve looks a bit less sloping than the blue one, \textit{i.e.},
the initial jump of kinetic energy (presumably caused by the ``virialization'')
for $ N_{\rm tot} = 1000 $ is more pronounced, but the subsequent increase
(associated with the multi-particle recombination) is slower.
The corresponding difference can be up to approximately 5\%.
It is difficult to say if this is a real physical effect, because our
algorithm of integration turned out to be less efficient for the completely
random distribution, so that the total accumulated error (estimated by
the conservation of energy) often was also about 5\%.
Anyway, the above-mentioned discrepancy in the average curves is much less
than the r.m.s.\ deviations $ \hat{\sigma}_t $ and $ \hat{\sigma}_{\rm ver} $
presented in Table~\ref{tab:Resulting_temperature} and
Fig.~\ref{fig:Resulting_temperature}.

In summary, one can conclude that a reasonable way to reduce the computational
cost is to perform simulations with a less number of test particles but
for a larger set of initial conditions.
Then, a greater variation of the individual curves should be quickly
compensated by averaging over a larger statistical sample, and the resulting
average curve will be sufficiently accurate.
For example, the cost of 15~simulations with 125~particles is 20~times cheaper
than the cost of 5~simulations with 1000~particles, while the resulting
average curves coincide with each other almost perfectly.

\subsection*{References:}


%

\end{document}